\documentclass[cits]{PoS}\usepackage{amsmath,cite} 
\title{Precise Ratios of Decay Constants of Vector over
Pseudoscalar $B_{(s)}$ Mesons}\ShortTitle{Precise Ratios of Decay
Constants of Vector over Pseudoscalar $B_{(s)}$ Mesons}
\author{Wolfgang Lucha\\Institute for High Energy Physics,
Austrian Academy of Sciences, Nikolsdorfergasse 18, A-1050 Vienna,
Austria\\E-mail: \email{Wolfgang.Lucha@oeaw.ac.at}}
\author{\speaker{Dmitri Melikhov}\\D.~V.~Skobeltsyn Institute of
Nuclear Physics, Moscow State University, 119991, Moscow, Russia,
and\\Faculty of Physics, University of Vienna, Boltzmanngasse 5,
A-1090 Vienna, Austria\\E-mail: \email{dmitri\_melikhov@gmx.de}}
\author{Silvano Simula\\INFN, Sezione di Roma Tre, Via della Vasca
Navale 84, I-00146 Roma, Italy\\E-mail:
\email{simula@roma3.infn.it}}

\abstract{The relative magnitude of the decay constants of the
pseudoscalar and vector beauty mesons poses (in contrast to the
case of charmed mesons) a long-standing puzzle. We revisit this
problem within the framework of our recent improvements of the QCD
sum-rule formalism for relating observable properties of mesons to
quantum chromodynamics and are led to conclude that the decay
constants of the beauty vector mesons are undoubtedly smaller than
those of their pseudoscalar counterparts.}

\FullConference{The European Physical Society Conference on High
Energy Physics\\22--29 July 2015\\Vienna, Austria}

\begin{document}\section{Technical keystone of motivation:
\emph{The advantage of being \emph{ratio\/}nal}\/}By chance,
uncertainties of physical observables (\emph{e.g.}, decay
constants) may partially cancel in ratios of these quantities.
This may render predictions for such ratios more precise than
those~for the individual quantities. We intend to exploit this
serendipity to relate the $B_{(s)}^{(*)}$-meson decay constants.

\section{QCD sum rules for beauty vector mesons}We analyze the
decay constants $f_V$ of the beauty \emph{vector\/} mesons $B^*$
and $B_s^*$ --- defined for vector mesons $V$ with mass $M_V$ and
polarization vector $\varepsilon_\mu(p)$ in terms of heavy--light
quark vector~currents $j_\mu(x)=\bar q(x)\,\gamma_\mu\,Q(x)$
according to
$\langle0|j_\mu(0)|V(p)\rangle=f_V\,M_V\,\varepsilon_\mu(p)$ ---
by means of QCD sum rules;~in particular, we'll be interested in
the \emph{relative\/} magnitude of the decay constants of such
vector mesons and their pseudoscalar counterparts, as (in contrast
to initial belief and the charmed-meson case)~our preliminary
results \cite{LMSFH} provided a first hint that in the beauty
sector the decay constants of the vector mesons are
\emph{smaller\/} than those ($f_P$) of the pseudoscalar ones.
Starting from the two-point~correlators$${\rm i}\int{\rm
d}^4x\,{\rm e}^{{\rm i}\,p\,x}\,\langle0|T\!\left(j_\mu(x)\,
j^\dagger_\nu(0)\right)\!|0\rangle=\left(-g_{\mu\nu}+\frac
{p_\mu\,p_\nu}{p^2}\right)\Pi(p^2)+\frac{p_\mu\,p_\nu}{p^2}\,
\Pi_L(p^2)$$ subjected to \emph{operator product expansion\/}
(OPE), Borel transformation to a Borel variable, $\tau,$ and the
postulate that above \emph{effective thresholds\/} $s_{\rm
eff}(\tau)$ all unknown contributions of excited and continuum
hadron states equal those of perturbative QCD, we eventually
deduce from $\Pi(p^2)$ the QCD sum~rule$$f_V^2\,M_V^2\,{\rm
e}^{-M_V^2\,\tau}=\hspace{-2ex}\int\limits_{(m_Q+m)^2}^{s_{\rm
eff}(\tau)}\hspace{-2ex}{\rm d}s\,{\rm e}^{-s\,\tau}\,\rho_{\rm
pert}(s,\mu)+\Pi_{\rm power}(\tau,\mu)\equiv\Pi_{\rm
dual}(\tau,s_{\rm eff}(\tau))\ .$$The right-hand side of this
relation forms the ``dual correlator'' $\Pi_{\rm dual}(\tau,s_{\rm
eff}(\tau)),$ which receives both perturbative contributions
usually encoded in a dispersion integral of an appropriate
spectral~density$$\rho_{\rm pert}(s)=\rho^{(0)}(s,m_b)
+\frac{\alpha_{\rm s}(\nu)}{\pi}\,\rho^{(1)}(s,m_b)
+\frac{\alpha^2_{\rm s}(\nu)}{\pi^2}\,\rho^{(2)}(s,m_b,\mu)
+\cdots$$and ``power'' contributions involving the vacuum
condensates that parameterize all non-perturbative effects, and
which fixes our predictions for \emph{dual\/} mass and decay
constant of the hadron under study:$$M_{\rm dual}^2(\tau)
\equiv-\frac{{\rm d}}{{\rm d}\tau}\log\Pi_{\rm dual}(\tau,s_{\rm
eff}(\tau))\ ,\qquad f_{\rm dual}^2(\tau)\equiv\frac{{\rm
e}^{M_V^2\,\tau}}{M_V^2}\,\Pi_{\rm dual}(\tau,s_{\rm eff}(\tau))\
.$$ The effective threshold $s_{\rm eff}(\tau)$ is determined by
minimizing, for polynomial \emph{Ans\"atze\/} of low orders~$n$
$$s^{(n)}_{\rm eff}(\tau)=\sum_{j=0}^ns^{(n)}_j\,\tau^j$$with
expansion coefficients $s^{(n)}_j,$ the deviation of this
predicted meson mass from its measured~value
$$\chi^2\equiv\frac{1}{N}\sum_{i=1}^N\left[M^2_{\rm
dual}(\tau_i)-M_V^2\right]^2$$over a set of $N$ equidistant
discrete points $\tau_i$ in the admissible region of $\tau$
\cite{LMSET}. Our results'~spread~for polynomial orders $n=1,2,3$
enables us to estimate the \emph{systematic\/} uncertainties
inherent to the~QCD sum-rule formalism \cite{LMSAU}. Our novel
ideas met great success when being applied to heavy
mesons~\cite{LMSD*Bs}. The actual application of this approach
requires obvious numerical ingredients, collected in
Table~\ref{Tab:NIP}.

\begin{table}[ht]\begin{center}\caption{Numerical parameter values
required as input to the operator product expansions for beauty
mesons.}\label{Tab:NIP}\vspace{2ex}
\begin{tabular}{lcc}\hline\hline&\\[-2.7ex]
\multicolumn{2}{c}{Quantity}&Numerical input value\\[.5ex]\hline
\\[-2.7ex]Light-quark $\overline{\rm MS}$ mass&
$\overline{m}(2\;\mbox{GeV})$&$(3.42\pm0.09)\;\mbox{MeV}$
\\Strange-quark $\overline{\rm MS}$ mass&
$\overline{m}_s(2\;\mbox{GeV})$&$(93.8\pm2.4)\;\mbox{MeV}$
\\Bottom-quark $\overline{\rm MS}$ mass&
$m_b\equiv\overline{m}_b(\overline{m}_b)$&$(4247\pm34)\;\mbox{MeV}$
\cite{LMSmb}\\Strong coupling& $\alpha_{\rm
s}(M_Z)$&$0.1184\pm0.0020$\\Light-quark condensate&$\langle\bar
q\,q\rangle\equiv\langle\bar
q\,q\rangle(2\;\mbox{GeV})$&$-[(267\pm17)\;\mbox{MeV}]^3$\\
Strange-quark condensate& $\langle\bar
s\,s\rangle\equiv\langle\bar s\,s\rangle(2\;\mbox{GeV})$&
$(0.8\pm0.3)\times\langle\bar
q\,q\rangle(2\;\mbox{GeV})$\\Two-gluon condensate&$\displaystyle
\left\langle\frac{\alpha_{\rm s}}{\pi}\,G\,G\right\rangle$&
$(0.024\pm0.012)\;\mbox{GeV}^4$\\[1ex]\hline\hline\end{tabular}
\end{center}\end{table}

\section{Decay-constant ratio of $B^*$ and $B$ mesons}
\begin{figure}[hbt]\begin{center}
\includegraphics[scale=.48]{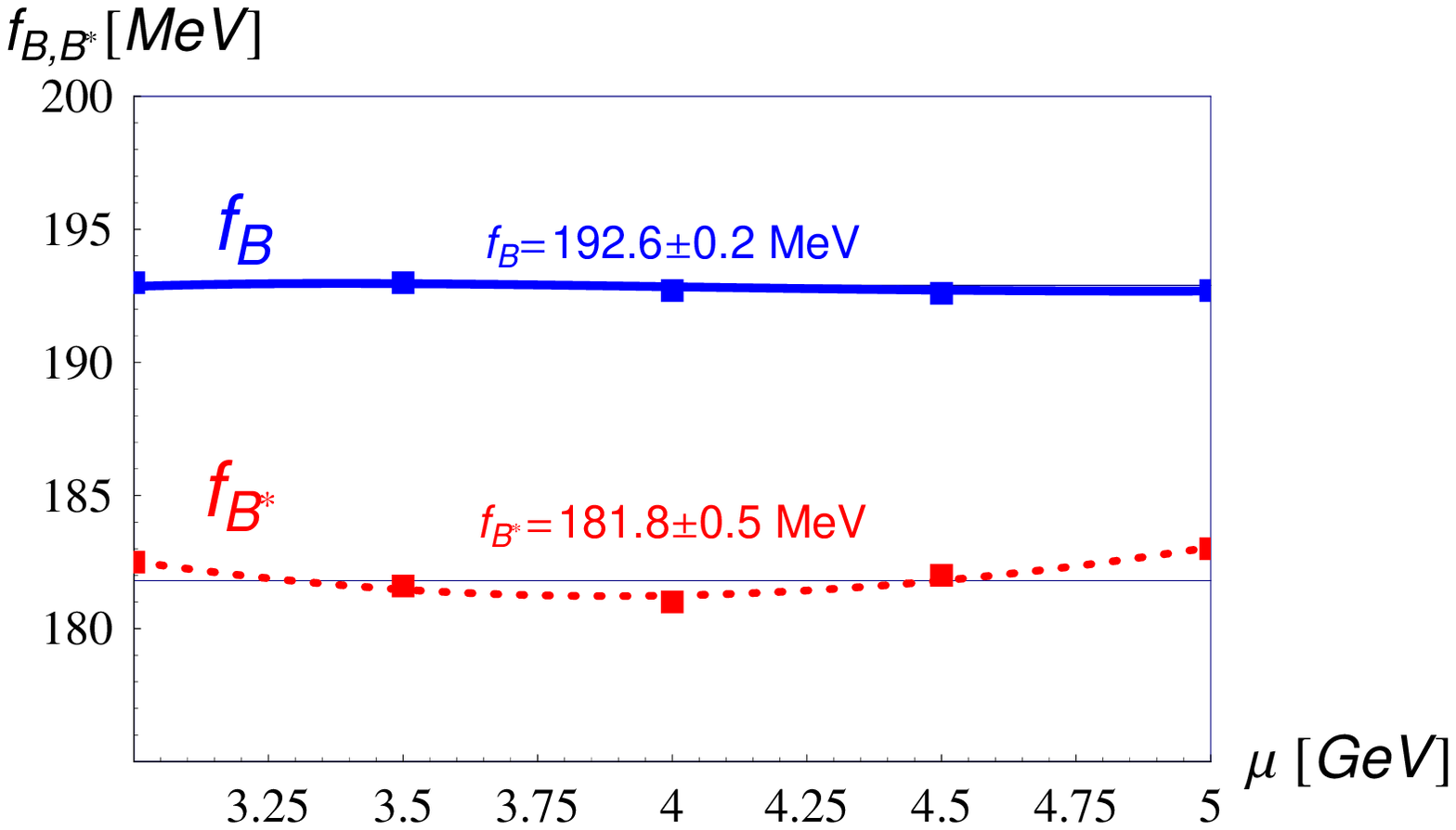}\\
\includegraphics[scale=.48]{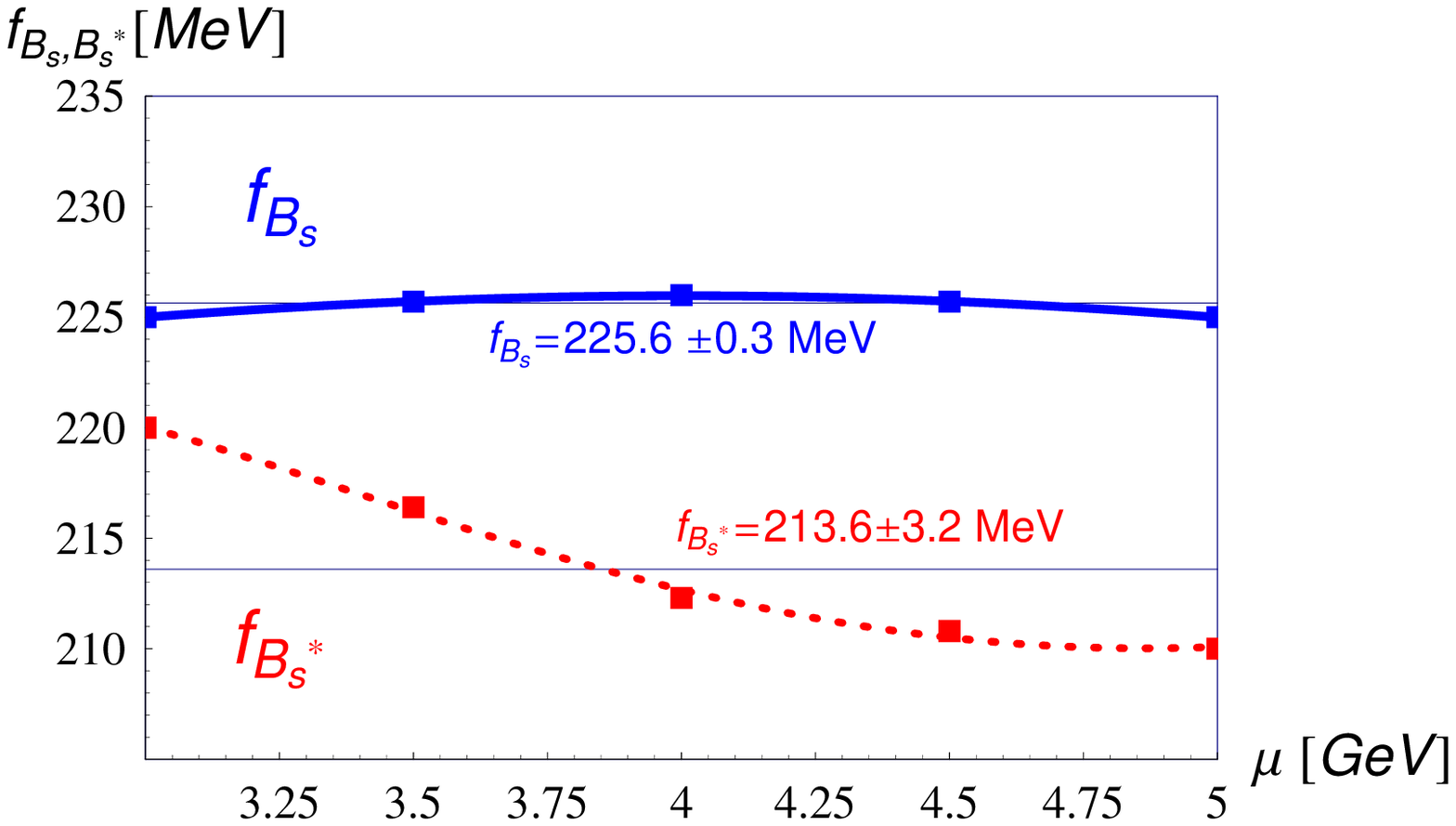}
\caption{$B^{(*)}$ (top) and $B_s^{(*)}$ (bottom) meson decay
constants $f_{(s)}^{(*)}$ as function of the renormalization
scale~$\mu.$}\label{Fig:BDC}\end{center}\end{figure}

\begin{figure}[hbt]\begin{center}
\includegraphics[scale=.375]{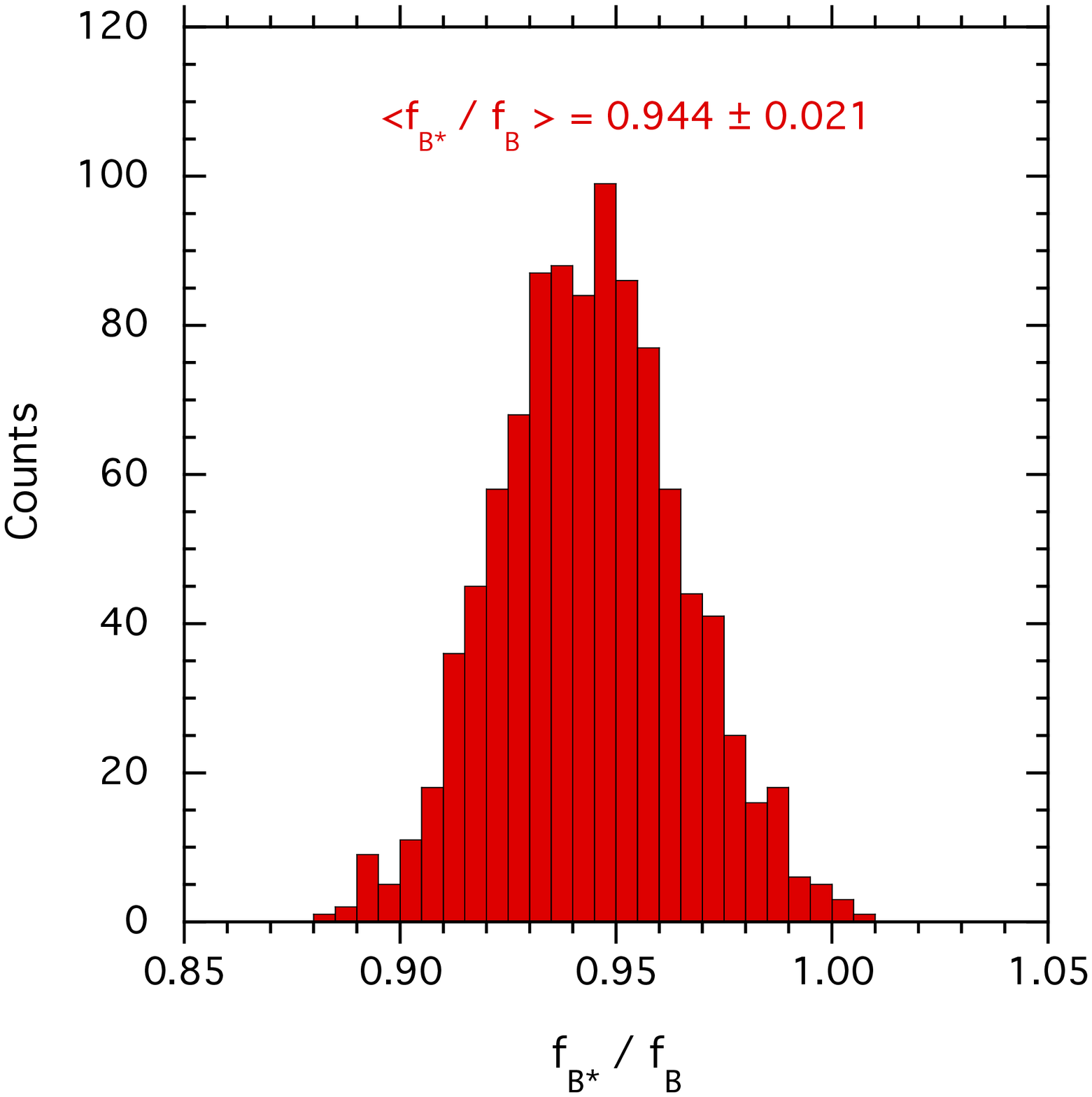}\\
\includegraphics[scale=.375]{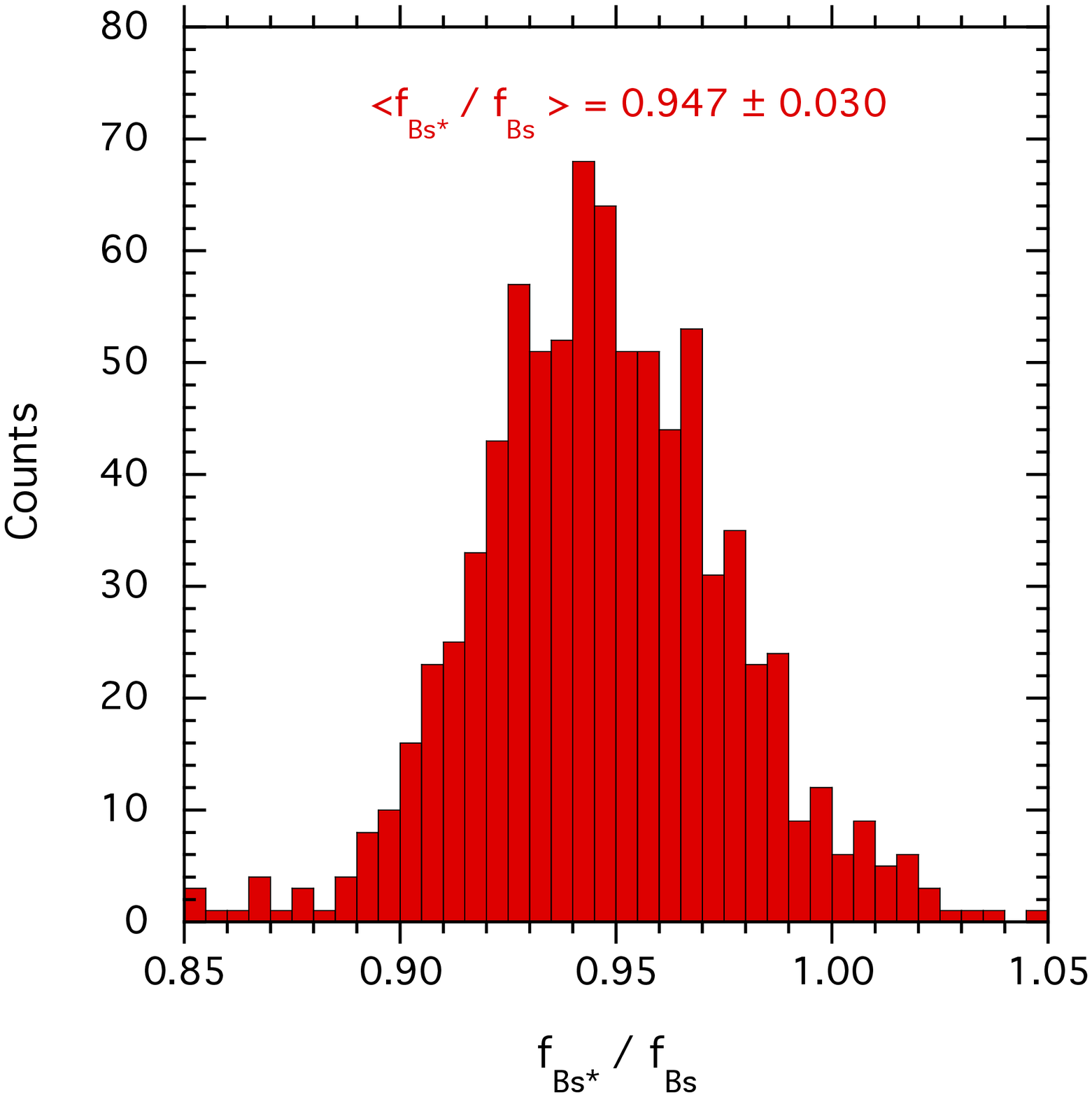}
\caption{Gaussian-like distributions of vector vs.\ pseudoscalar
decay-constant ratios for nonstrange,~$f_{B^*}/f_B$ (top), and
strange, $f_{B_s^*}/f_{B_s}$ (bottom), beauty mesons from a
bootstrap study relying on 1000 generated~events.}\label{Fig:DCR}
\end{center}\end{figure}

Within the
advanced formalism constructed and corroborated in a sequence of
papers \cite{LMSAU,LMSET,LMSD*Bs},~the extraction of the decay
constants from QCD sum rules proceeds along meanwhile well-paved
paths:\begin{itemize}\item So far, our perturbative spectral
density $\rho_{\rm pert}(s)$ has been derived, in terms of the
heavy-quark pole mass, up to order $O(\alpha_{\rm s}^2)$ or,
equivalently, up to three-loop accuracy \cite{SD}. A
reorganization of this perturbative expansion in terms of the
$\overline{\rm MS}$ mass of the bottom quark bears the potential
to improve the obviously confidence-inspiring \emph{hierarchy\/}
of the perturbative contributions \cite{JL}.\item The unavoidable
\emph{truncation\/} of the (perturbative) spectral densities and
the (non-perturbative) power contributions spoils the independence
of QCD sum-rule extractions of any observables from the
renormalization scale $\mu$ and provokes their (unphysical) $\mu$
dependence. Perturbative convergence and reproducibility of the
$B^*$-meson's mass confine the acceptable values of $\mu$~to
$$3\;\mbox{GeV}\le\mu\le5\;\mbox{GeV}\ .$$\item The allowed range
of the \emph{Borel variable\/} $\tau$ is defined by requiring the
$B$- and $B^*$-meson~masses and the $B$--$B^*$ mass splitting to
be predictable with an error less than $5\;\mbox{MeV}$ over this
$\tau$ region:$$0.01\;\mbox{GeV}^{-2}\le\tau\le0.31\;
\mbox{GeV}^{-2}-0.05\,\mu\;\mbox{GeV}^{-3}\ .$$\end{itemize}In
addition to the systematic errors, roughly measured by our
algorithm for extracting an observable from a QCD sum rule
\cite{LMSAU,LMSET}, the limited precision of the input parameter
values induces OPE-related \emph{statistical\/} uncertainties. Our
findings for $f_{B^*}$ exhibit a linear dependence on the relevant
OPE input,\begin{align*}f_{B^*}^{\rm dual}(m_b,\langle\bar
q\,q\rangle,\langle\mbox{$\frac{\alpha_{\rm
s}}{\pi}$}\,G\,G\rangle)=(181.8\pm4_{\rm syst})&\times
\left(1-\frac{11}{181.8}\,\frac{m_b-4.247\;\mbox{GeV}}{0.034\;\mbox{GeV}}
\right)\\\times\left(1+\frac{7}{181.8}\,\frac{|\langle\bar
q\,q\rangle|^{1/3}-0.267\;\mbox{GeV}}{0.017\;\mbox{GeV}}\right)&\times
\left(1-\frac{1}{181.8}\,\frac{\langle\mbox{$\frac{\alpha_{\rm
s}}{\pi}$}\,G\,G\rangle-0.024\;\mbox{GeV}^4}{0.012\;{\rm
GeV}^4}\right)\mbox{MeV}\ ,\end{align*}but insensitivity to the
renormalization scale in its range (Fig.~\ref{Fig:BDC}). Averaging
over assumed Gaussian distributions of all the OPE parameters but
a flat distribution of the scale $\mu$ eventually yields
(Fig.~\ref{Fig:DCR})$$f_{B^*}=\left(181.8\pm13.1_{\rm
OPE}\pm4_{\rm syst}\right)\mbox{MeV}\ .$$\newpage

\noindent The \emph{ratios\/} of decay constants benefit from huge
\emph{cancellations\/} among their OPE uncertainties. Their
remaining OPE errors arise primarily from the gluon condensate and
their total errors are~dominated by the \emph{systematic\/}
uncertainties. Confronting $f_{B^*}$ with our earlier finding
\cite{LMSmb} for the $B$-meson's decay constant $f_B$, we find the
$B^*$-meson's decay constant to lie $2.7\sigma$ \emph{below\/} the
$B$-meson's one \cite{LMSDCR}~(Fig.~\ref{Fig:DCR}):
$$\frac{f_{B^*}}{f_B}=0.944\pm0.011_{\rm OPE}\pm0.018_{\rm syst}
=0.944\pm0.021\;\lneqq1\ .$$

\section{Decay-constant ratio of $B_s^*$ and $B_s$ mesons}Without
going into details, let us now mimic the treatment of nonstrange
beauty mesons~for~the case of strange beauty mesons, stressing the
unequal aspects. The response of $f_{B^*_s}$ to $1\sigma$
variations~is\begin{align*}f_{B^*_s}^{\rm
dual}(\mu=\overline{\mu},m_b,\langle\bar
s\,s\rangle,\langle\mbox{$\frac{\alpha_{\rm
s}}{\pi}$}\,G\,G\rangle) =(213.6\pm6_{\rm
syst})&\times\left(1-\frac{13.2}{213.6}\,
\frac{m_b-4.247\;\mbox{GeV}}{0.034\;\mbox{GeV}}\right)\\\times
\left(1+\frac{11.8}{213.6}\,\frac{|\langle\bar
s\,s\rangle|^{1/3}-0.248\;\mbox{GeV}}{0.033\;\mbox{GeV}}\right)&\times
\left(1-\frac{1}{213.6}\,\frac{\langle \mbox{$\frac{\alpha_{\rm
s}}{\pi}$}\,G\,G\rangle-0.024\;\mbox{GeV}^4}{0.012\;\mbox{GeV}^4}\right)
\mbox{MeV}\ .\end{align*}Unlike the $B^*$ meson, the $B_s^*$ meson
exhibits a pronounced dependence on the renormalization~scale,
$$f_{B_s^*}^{\rm dual}(\mu)=213.6\;\mbox{MeV}
\left(1-0.12\log\frac{\mu}{\overline{\mu}}+0.11\log^2\frac{\mu}
{\overline{\mu}}+0.43\log^3\frac{\mu}{\overline{\mu}}\right),$$introducing
a kind of \emph{average\/} $\overline{\mu}$ of the renormalization
scale: $\overline{\mu}=3.86\;\mbox{GeV}.$ The resulting
$f_{B_s^*}$~reads$$f_{B_s^*}=(213.6\pm18.2_{\rm OPE}\pm6_{\rm
syst})\;\mbox{MeV}\ .$$\newpage For completeness, the
corresponding relations of the pseudoscalar strange beauty meson
$B_s$~are\begin{align*}f_{B_s}^{\rm dual}(m_b,\langle\bar
s\,s\rangle,\langle\mbox{$\frac{\alpha_{\rm
s}}{\pi}$}\,G\,G\rangle) =(225.6\pm3_{\rm
syst})&\times\left(1-\frac{14.1}{225.6}\,
\frac{m_b-4.247\;\mbox{GeV}} {0.034\;\mbox{GeV}}\right)\\\times
\left(1+\frac{11.5}{225.6}\,\frac{|\langle\bar
s\,s\rangle|^{1/3}-0.248\;\mbox{GeV}}{0.033\;\mbox{GeV}}\right)&\times
\left(1+\frac{1}{225.6}\,\frac{\langle\mbox{$\frac{\alpha_{\rm s}}
{\pi}$}\,G\,G\rangle-0.024\;\mbox{GeV}^4}{0.012\;\mbox{GeV}^4}\right)
\mbox{MeV}\end{align*}for the behaviour of $f_{B_s}$ under
$1\sigma$ variations of all crucial OPE parameters and, as our
$f_{B_s}$ prediction,$$f_{B_s}=(225.6\pm18.3_{\rm OPE}\pm3_{\rm
syst})\;\mbox{MeV}\ .$$In the case of strange beauty mesons, their
decay-constant ratio is thus $1.7\sigma$ \emph{below} unity
\cite{LMSDCR} (Fig.~\ref{Fig:DCR}):$$\frac{f_{B_s^*}}{f_{B_s}}=
0.947\pm0.023_{\rm OPE}\pm0.020_{\rm syst}=0.947\pm0.030\;\lneqq1\
.$$

\section{Observations, conclusions, and comparison}Our QCD sum-rule
analysis of the decay constants of $B_{(s)}^{(*)}$ mesons
\cite{LMSDCR} provides a lot of insights:\begin{itemize}\item As
in the case of the pseudoscalar heavy mesons \cite{LMSD*Bs}, the
highly unsatisfactory convergence of the perturbative expansion
formulated in terms of the pole mass of the heavy quark found also
for vector heavy mesons enforces conversion of the OPE to the
$\overline{\rm MS}$ quark-mass definition \cite{LMSDCR}.\item
Acceptable reproduction of the experimentally measured masses of
the beauty vector mesons by our QCD sum-rule dual predictions
necessitates a correlation between the upper boundary of the
adoptable Borel-variable range and the renormalization scale
chosen for evaluation \cite{LMSDCR}.\item Very accurate
reproduction of the meson masses and their splitting, enabled by
our concept of extraction of an observable \cite{LMSET}, is
imperative for the smallness of the systematic uncertainties.\item
A study of beauty-meson decay constants within the realm of
lattice-regularized QCD carried out practically simultaneously to
our analysis gets \cite{HPQCD15}, in perfect agreement with our
findings,$$\frac{f_{B^*}}{f_B}=0.941\pm0.026\ ,\qquad
\frac{f_{B_s^*}}{f_{B_s}}=0.953\pm0.023\ .$$\end{itemize}So, the
outcomes of the present study add a great deal of credibility to
our initial observation~\cite{LMSFH}:~the decay constants of
vector beauty mesons are, beyond doubt, smaller than those of
their pseudoscalar counterparts; hence, we are no longer stunned
by our inability to reproduce the claims of Refs.~\cite{DCR>1}.

\end{document}